\documentclass{article}
\usepackage{spconf,amsmath,graphicx}
\usepackage{mathtools}
\usepackage{bm}

\title{Attention mechanism in speaker recognition: what does it learn in deep speaker embedding? }
\name{Qiongqiong Wang, Koji Okabe, Kong Aik Lee, Hitoshi Yamamoto, Takafumi Koshinaka}
\address{Biometrics Research Laboratories, NEC Corporation, Japan}
\begin{document}
%
\maketitle
\begin{abstract}
This paper presents an experimental study on deep speaker embedding with an attention mechanism that has been found to be a powerful representation learning technique in speaker recognition.  
In this framework, an attention model works as a frame selector that computes an attention weight for each frame-level feature vector, in accord with which an utterance-level representation is produced at the pooling layer in a speaker embedding network. 
In general, an attention model is trained together with the speaker embedding network on a single objective function, and thus those two components are tightly bound to one another.  
In this paper, we consider the possibility that the attention model might be decoupled from its parent network 
and assist other speaker embedding networks and even conventional i-vector extractors.  
This possibility is demonstrated through a series of experiments on a NIST Speaker Recognition Evaluation (SRE) task, 
with 9.0\% EER reduction and 3.8\% $\rm min$$\rm C_{primary}$ reduction when the attention weights are applied to i-vector extraction.  Another experiment shows that DNN-based soft voice activity detection (VAD) can be effectively combined with the attention mechanism to yield further reduction of $\rm min$$\rm C_{primary}$ by 6.6\% and 1.6\% in deep speaker embedding and i-vector systems, respectively.


\end{abstract}
\section{Introduction}
\label{sec:intro}

With the recent success of deep learning over a wide range of machine learning tasks, including automatic speech recognition (ASR) \cite{Yu2012}\cite{Hinton2012} and facial recognition \cite{SchroffKP15}, 
investigation of the use of deep neural networks (DNNs) for speaker recognition has increased dramatically,  
including the possible use of  DNNs to replace conventional feature extraction \cite{lei2014novel}\cite{mclaren2015advances}, back-end modeling \cite{Chien2017}, 
and the entire front-to-back-end processing pipeline in an end-to-end manner \cite{snyder2016}\cite{li2017deep}.

On the basis of the i-vector paradigm \cite{dehak2011front}, it was shown in \cite{lei2014novel}\cite{mclaren2015advances} that deep neural networks derived from acoustic models in ASR 
can be used as universal background models (UBMs) to provide phoneme posteriors as well as bottleneck features.
While these have shown better performance than conventional UBMs based on Gaussian mixture models (GMMs), they
have the drawback of language dependency \cite{zheng2015exploring} and also require expensive phonetic transcriptions for training \cite{tian2016improving}.
%

More recently, DNNs have been shown to be useful for extracting speaker-discriminative feature vectors independently from the i-vector framework. 
With the help of a certain amount of training data, such approaches lead to better results, particularly under conditions of short-duration utterances.
It was shown in \cite{nagrani2017voxceleb}\cite{li2017deep} that DNNs achieve better accuracy than do i-vectors. 
In \cite{snyder2017deep}, statistics pooling was employed to aggregate frame-level speaker representations to obtain an utterance-level representation, i.e.,
speaker embedding (x-vector), with a fixed number of dimensions, regardless of the length of the input utterance.

Most recent studies conducted from a different perspective \cite{bhattacharya2017deep}\cite{chowdhury2017attention}\cite{okabe}\cite{zhu2018} have incorporated attention mechanisms \cite{raffel2015feed}, 
which have produced significant improvements in machine translation \cite{bahdanau2015}. 
In the scenario of speaker recognition, an importance measure is computed by means of a small attention network that works as a part of the speaker embedding network,
as well as in the pooling layer utilized for calculating the weighted mean of frame-level feature vectors.
It has been applied to text-dependent \cite{chowdhury2017attention} and text-independent speaker recognition, including fixed-duration \cite{bhattacharya2017deep} and variable-duration \cite{okabe}\cite{zhu2018} settings.

The attention mechanism is a powerful technique which offers a way to obtain an even more discriminative utterance-level representation by explicitly selecting frame-level features that better represent speaker characteristics. 
Its remarkable advantage is that the attention model is automatically trained as a part of the deep speaker embedding network, in accord with a single objective function.  
Without attaching any additional labels, such as which frames are important, the attention model is optimized together with its parent network just so as to minimize speaker identification errors.
This configuration of the attention mechanism suggests that the weight computed by the attention model is tightly bound to the frame-level features produced by the speaker embedding network. 

 Thus, a question arises: What does the attention model actually learn?  It may not learn a general importance of frames but, rather, learn something specific to frame-level features which the coupled speaker embedding network produces.  
 Even if it works well with the coupled speaker embedding network, it may not work well with other networks or conventional i-vector extractors, for which the i-vector paradigm continues to have its own advantages under some practical conditions, including relatively long speech [13].  
 This paper attempts to give an answer to the above question through a series of experiments and demonstrates that the attention model coupled with a deep speaker embedding network can work with other networks and even with i-vector extractors. 
 To the best of our knowledge, no study has yet been done on deep speaker embedding from such a perspective.

The remainder of this paper is organized as follows: 
Section 2 describes a conventional method for extracting deep speaker embedding with an attention mechanism. 
Section 3 presents fundamental formulae for the i-vector framework and how the attention weights from a deep speaker embedding network can be applied to it. 
The experimental setup and results are presented in Section 4. 
Section 5 summarizes our work.


\vspace*{-2mm}
\section{Deep speaker embedding}
\label{sec:format}
Speaker embeddings are low-dimensional representations of speech utterances with the property of capturing speaker characteristic in recognition tasks \cite{Niko2018}.
Presented below is a brief description of deep speaker embedding obtained with a DNN having either a non-attentive or an attentive statistics pooling layer. 
\vspace*{-3mm}
\subsection{Embedding via statistics pooling}

\begin{figure}[t]
\centering
\includegraphics[width=\linewidth]{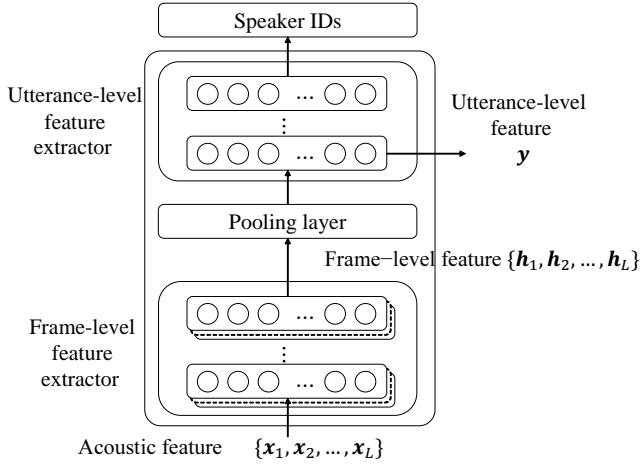}
\vspace*{-5mm}
\caption{Conventional DNNs for extracting utterance-level speaker representations \cite{snyder2017deep}}
\label{fig:dnn}
\vspace*{-3mm}
\end{figure}

A conventional DNN for extracting an utterance-level speaker representation consists of three modules, as shown in Figure~\ref{fig:dnn}.
The first module is a frame-level feature extractor. 
The input to this module is a sequence of acoustic features, e.g., Mel-frequency Cepstral Coefficients (MFCCs) and filter-bank coefficients. 
After considering relatively short-term acoustic features, this module outputs frame-level features.
Any type of neural networks is applicable for the extractor, e.g., a Time-Delay Neural Network (TDNN) \cite{snyder2017deep}, Convolutional Neural Network (CNN) \cite{nagrani2017voxceleb}\cite{li2017deep}, LSTM \cite{bhattacharya2017deep}\cite{chowdhury2017attention}, or Gated Recurrent Unit (GRU) \cite{li2017deep}.

The second module is a pooling layer that converts variable-length frame-level features into a fixed-dimensional vector. 
The most standard type of pooling layer obtains the mean vector $\bm{\mu}$ of all frame-level features $\bm{h}_t(t=1, \cdots,L)$:
\begin{equation}
\bm{\mu}=\frac{1}{L}\sum_{t=1}^L \bm{h}_t,\label{eq:nonattmean}
\end{equation}
where $L$ indicates the number of frames.

In \cite{snyder2017deep}\cite{okabe}, the second-order statistics, the standard deviation vector, $\bm{\sigma}$ was used as well:
\begin{equation}
\bm{\sigma}=\sqrt{\frac{1}{L}\sum_{t=1}^L \bm{h}_t\odot \bm{h}_t-\bm{\mu} \odot \bm{\mu}}, \label{eq:nonattsdv}
\end{equation}
where $\odot$ represents the Hadamard product. 

The third module produces utterance-level representations for which a number of fully-connected hidden layers are stacked.
One of these hidden layers is often designed to have a smaller number of units (i.e., to be a bottleneck layer), which forces
the information brought from the preceding layer into a low dimensional representation. 
The output is a softmax layer, with each of its output nodes corresponds to one speaker ID.
For training, we employ back-propagation with cross-entropy loss. 
We can then use bottleneck features in the third module as utterance-level representations. 
Some studies refrain from using softmax layers and achieve end-to-end neural networks by using contrastive loss \cite{nagrani2017voxceleb} or triplet loss \cite{li2017deep}. 
Probabilistic linear discriminant analysis (PLDA) \cite{ioffe2006probabilistic}\cite{prince2007probabilistic} can also be used for measuring the distance between two utterances \cite{snyder2017deep}\cite{bhattacharya2017deep}.

\subsection{ Attentive speaker embedding}
\label{ssec:att}
It is often the case that frame-level features of some frames are more unique and important for discriminating speakers than others in a given utterance. 
Recent studies \cite{bhattacharya2017deep}\cite{chowdhury2017attention}\cite{okabe}\cite{zhu2018} have applied attention mechanisms to speaker recognition for the purpose of frame selection by automatically calculating the importance of each frame.

\begin{figure}[t]
\centering
\includegraphics[width=\linewidth]{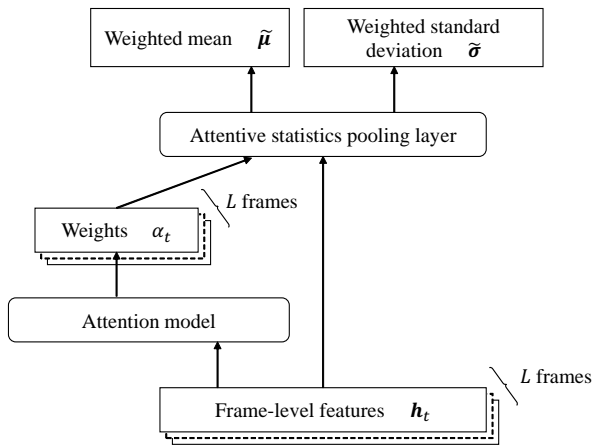}
\vspace*{-5mm}
\caption{Attentive statistics pooling \cite{okabe}}
\label{fig:attentive_statistics_pooling}
\end{figure}

As shown in Figure~\ref{fig:attentive_statistics_pooling}, an attention model works in conjunction with the original DNN and calculates a scalar score $e_t$ for each frame-level feature
\begin{equation}
e_t=\bm{v}^Tf(\bm{Wh}_t+\bm{b})+k, \label{eq:et}
\end{equation}
where $f(\cdot)$ is a non-linear activation function, such as a tanh or ReLU function. 
The score is normalized over all frames by a softmax function so as to add up to the following unity:
\begin{equation}
\alpha_t=\frac{\exp(e_t)}{\sum_{\tau=1}^L \exp(e_\tau)}. \label{eq:att}
\end{equation}
The normalized score $\alpha_t$ is then used as the weight in the pooling layer to calculate the following weighted mean and standard deviation vectors: 
\begin{equation}
\tilde{\bm{\mu}}=\sum_{t=1}^L \alpha_t \bm{h}_t, \label{eq:attmean}
\end{equation}
\begin{equation}
\tilde{\bm{\sigma}}=\sqrt{\sum_{t=1}^L \alpha_t \bm{h}_t\odot \bm{h}_t-\tilde{\bm{\mu}} \odot \tilde{\bm{\mu}}}. \label{eq:attsdv}
\end{equation}
In this way, the utterance-level representations in the form of weighted statistics focus on important frames and hence
become more speaker discriminative.
Notice that, if we set $\alpha_t=\frac{1}{L}$, attentive pooling in Eqs.~\eqref{eq:attmean} and \eqref{eq:attsdv} falls back to the non-attentive (i.e., equal weight) pooling in Eqs.~\eqref{eq:nonattmean} and \eqref{eq:nonattsdv}. 
Results in \cite{okabe} have shown considerable performance improvement in speaker verification tasks using attentive weights produced by an attention model in the pooling layer.

%

\section{WHAT AN ATTENTION MODEL LEARNS}
\label{sec:pagestyle}

\subsection{Motivation and objectives}

In the attentive speaker embedding described in Subsection~\ref{ssec:att}, the attention model is trained as a part of the speaker embedding network in accord with a single objective function so as to maximize the speaker-discriminative power.  
It is reasonable to assume that the weights produced by the attention model are tightly bound to the frame-level features which the speaker embedding network produces. 
The question is whether the weights and frame-level features are still able to play their roles when they are decoupled from the DNN. 
To the best of our knowledge, there have been few studies on such a perspective toward better understanding of deep speaker embedding.

\begin{figure}[h]
\centering
\includegraphics[width=0.4\textwidth]{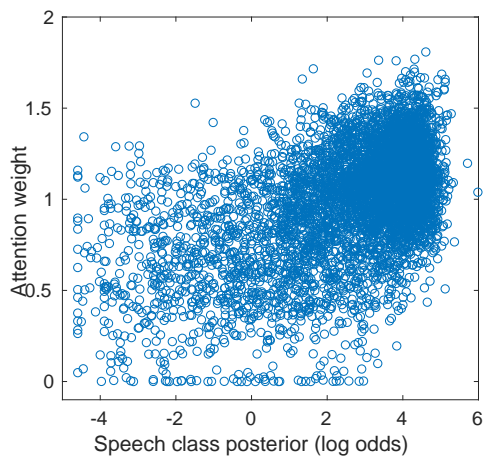}
\caption{Correlation between attention weights and speech class posteriors (log odds)}
\label{fig:plot}
\end{figure}

The use of an attention mechanism in deep speaker embedding (x-vector) extraction is largely driven by the statistics pooling layer. 
The aim is to find an optimal set of weights for each utterance such that higher weights are assigned to frames which are more unique and important than others in the statistics pooling operation. 
It is intuitive to conjecture that frames receiving higher weights correspond to certain phonetic classes (e.g., vowels) which are more effective or useful to discriminating among speakers. 
Another line of thought has suggested that the attention weight might be associated with just simple speech versus non-speech classes. 
Figure ~\ref{fig:plot} shows a scatter plot for the attention weights and speech (versus non-speech) class posteriors (log odds) on the y- and x- axes, respectively, for one utterance drawn from the SRE’16 corpus \cite{sre}. The speech class posterior is estimated using an LSTM neural network, 
where the non-speech class encompasses laughter, unclear voices, noise-like (noise, sigh, lip smack, cough and breath) phenomena, and silence \cite{yamamoto}. 
See Section~\ref{sec:exp} for details regarding datasets and our experimental setup. 
A simple analysis gives a normalized correlation coefficient of 0.37.
The weak correlation suggests that the attention weights relate to more than just speech/non-speech detection. 
\vspace*{-4mm}

\vspace*{-2mm}
In this study, we set out to consider various aspects of the attention weights mentioned above.
We present the results of three evaluations: 
(1) Using attentive frame-level features alone without attention weights, 
(2) Applying attention weights to non-attentive frame-level features from another deep speaker embedding network, and 
(3) Applying attention weights to statistics for i-vector extraction.  
In addition, we also try (4) Combining soft voice activity detection (VAD) as another kind of attention mechanism for better speaker recognition accuracy.  
Among those, (3) is especially new since the i-vector framework is quite different from a deep speaker embedding framework, such as that with x-vectors.  Details in this regard are presented in the next subsection.

\subsection{I-vector extraction with attention weights}

\subsubsection{i-vector extraction}
\label{ssec:ivec}
The i-vector framework has been a standard in speaker recognition over the last decade \cite{dehak2011front}. 
In spite of the increasing research on DNN-based methods, the i-vector framework continues to have its own advantage for some conditions, including relatively long speech \cite{snyder2017deep}.

An i-vector is a low-dimensional vector in total variability space to which factor analysis allows the projection of an utterance \cite{dehak2011front}.
It is assumed that a GMM-supervector $\bm{M}$, corresponding to an utterance, can be modeled as
\begin{equation}
\bm{M}=\bm{m}+\bm{Tw},\label{eq:ivec}
\end{equation}
where $\bm{m}$ is the speaker- and channel-independent supervector typically taken from a universal background model (UBM),
the total variability matrix (TVM) $\bm{T}$ is a rectangular matrix of low rank.
An i-vector $\bm \phi$ is the posterior mean of the latent variable $\bm w$ in Eq.~\eqref{eq:ivec}.

The i-vector $\bm{\phi(\bm{x})}$ for a given utterance $\bm{x}$ can be obtained using the following equation:
\begin{equation}
\bm{\phi}(\bm{x})=(\bm{I}+\bm{T}^t\bm{\Sigma}^{-1} \bm{N}(\bm{x})\bm{T})^{-1}\bm{T}^t\bm{\Sigma}^{-1}\bm{F}(\bm{x}), \label{eq:wx}
\end{equation}
where $\bm{\Sigma}$ is the block-diagonal covariance matrix of supervectors obtained from the UBM. 
This equation uses two types of statistics w.r.t$.$ the utterance, $\bm{N(\bm{x})}$ and $\bm{F(\bm{x})}$. 
When a GMM is used as a UBM, for example, these statistics on mixture component $c$ of the UBM are written as follows:
\begin{equation}
\bm{N}_c(\bm{x})=\sum_{t=1}^L p(c|\bm{x}_t), \label{eq:Nc}
\end{equation}
\vspace*{-5mm}
\begin{equation}
\bm{F}_c(\bm{x})=\sum_{t=1}^L p(c|\bm{x}_t)(\bm{x}_t-\bm{\mu}_c),\label{eq:Fc}
\end{equation}
where $\bm{x}_t$ is the acoustic feature at the $t$-th frame of utterance $x$ with $L$ frames, $p(c|\cdot)$ corresponds to the posterior probability of mixture component $c$ for acoustic feature $\bm{x}_t$, 
and $\bm{\mu}_c$ is the mean of $c$.

\subsubsection{Extended i-vector extraction with attention weights}

As noted in Subsection~\ref{ssec:att}, it is often the case that some frames are more unique and important for discriminating speakers than others in a given utterance. 
In \cite{okabe}, it was shown that applying the attention model to an x-vector extraction network improves speaker verification performance, 
which indicates that the attention weights are able to represent the importance of deep frame-level features. 
Under the assumption that x-vectors are able to fairly represent a speech utterance, then the attention weights are supposed to be general in representing the importance of frames and independent from feature representation. 
In other words, the importance of frames is independent of the representations, i.e., deep speaker embedding (x-vector) \cite{snyder2017deep} or i-vector \cite{dehak2011front}.
For this reason, we propose application of the attention weights $\alpha_t$ trained with an x-vector network to i-vector extraction in order to emphasize more important frames. 

The attention weights $\alpha_ t$ in Eq.~\eqref{eq:att} can be seamlessly incorporated into a formulation in i-vector extraction \cite{yamamoto}. 
We extend the framework of standard i-vector extraction by incorporating attention weighs $\alpha_t$ into the statistics of Eqs.~\eqref{eq:Nc}$-$\eqref{eq:Fc} as
follows:
\begin{equation}
\bm{N}_c(\bm{x})=\sum_{t=1}^L (L\alpha_t)p(c|\bm{x}_t), \label{eq:newNc}
\end{equation}
\vspace*{-5mm}
\begin{equation}
\bm{F}_c(\bm{x})=\sum_{t=1}^L (L\alpha_t) p(c|\bm{x}_t)(\bm{x}_t-\bm{\mu}_c).\label{eq:newFc}
\end{equation}
The scale factor $L$ ensures that Eq.~\eqref{eq:wx} is kept the same for the new i-vector extraction.
Notice that Eqs.~\eqref{eq:newNc} and ~\eqref{eq:newFc} reduce to Eqs.~\eqref{eq:Nc} and ~\eqref{eq:Fc}, respectively, by using an equal weight $\alpha_t=\frac{1}{L}$ for all frames.

\section{Experiments}
\label{sec:exp}

We have evaluated the performance of speaker embedding on a speaker verification task in NIST 2016 Speaker Recognition Evaluation (SRE'16) \cite{sre}. 
In the experiments, we followed the fixed condition in which only the designated data are used for system training. 
We used English-language telephone recordings from SRE'04$-$'10, Switchboard, and Fisher for training of all our systems. 
The evaluation set consists of 1,986,728 trials taken from Call My Net telephone conversation spoken in Cantonese and Tagalog. 
%

In addition to equal error rate (EER), results are reported w.r.t$.$ the official performance metric for SRE'16, 
i.e., equalized $\rm C_{primary}$, the average detection cost function at two operating points \cite{sre}.
More precisely, we use the minimum cost ($\rm min$$\rm C_{primary}$) that indicates the best achievable performance without considering the issue of score calibration \cite{Niko2010}.

\subsection{Investigation of decoupled attention weights and frame-level features}
\label{ssec:expatt}
We first investigated decoupled attention weights and frame-level features extracted from a deep attentive speaker embedding network (x-vector).
\subsubsection{Experimental settings}
We used 20-dimensional MFCCs for every 10ms.
Sliding mean normalization with a 3-second window and energy-based voice activity detection (VAD) were then applied, in that order. 

The network structure, other than w.r.t$.$ its input dimensions, was exactly the same as the one shown in the recipe\footnote{egs/sre16/v2}
published in Kaldi's official repository \cite{povey2011kaldi}\cite{snyder2018vector}. 
A 5-layer TDNN with ReLU followed by batch normalization was used for extracting frame-level features. 
The number of hidden nodes in each hidden layer was 512. 
The dimension of a frame-level feature for pooling was 1500. 
Each frame-level feature was generated from a 15-frame context of acoustic feature vectors.

The pooling layer aggregates frame-level features to produce the mean and standard deviation, 
followed by 2 fully-connected layers with ReLU activation functions,
batch normalization, and a softmax output layer. 
The 512-dimensional bottleneck features from the first fully-connected layer were used as speaker embeddings.
We used ReLU followed by batch normalization for activation function $f (\cdot)$ in Eq.~\eqref{eq:et} of the attention model. 
The number of hidden nodes was 64.
\begin{table}[h] 
\centering
\begin{tabular}[t]{c|c|c|c}
\hline
System & Frame-level &Attention&Matched\\
&features & & or not\\
\hline \hline
S1 & Non-attentive & Non-attentive & Matched\\

S2 &Attentive & Attentive& Matched\\

S3 &Non-attentive &Attentive & Non-matched\\ 

S4 &Attentive &Non-attentive & Non-matched\\
\hline
\end{tabular}
\caption{Settings for systems S1$-$S4}
\label{tab:tab0}
\end{table}

We compared four systems with two pooling techniques to evaluate the coupled and decoupled attention weights and frame-level features, as shown in Table~\ref{tab:tab0}: 
(S1) frame-level features from a non-attentive network were aggregated without an attention mechanism,
(S2) frame-level features from an attentive network were aggregated with an attention mechanism,
(S3) attention weights in S2 was applied to frame-level features in S1, and 
(S4) frame-level features in S2 were used in non-attentive pooling. 
Note that S3 and S4 contain a mismatch between frame-level features and attention weights.

Mean subtraction, whitening, and length normalization \cite{garcia2011analysis} were applied to the speaker embedding as pre-processing steps
prior to PLDA scoring, and likelihood scores were then computed using a PLDA model with a speaker space of 512 dimensions.

\begin{table}[h] 
\centering
\begin{tabular}[t]{l||cc}
\hline
systems & EER(\%) &$\rm min$$\rm C_{primary}$ \\
\hline \hline
S1 (Matched) &11.47 &0.873 \\

S2 (Matched) &11.10 &0.853 \\

S3 (Non-matched) &11.06 &0.856 \\ 

S4 (Non-matched) &18.06 &0.996 \\
\hline
\end{tabular}
\caption{Performance of coupled and decoupled attention weights and attentive frame-level features}
\label{tab:tab1}
\end{table}

\subsubsection{Experimental results and analyses}

Experimental results w.r.t$.$ systems S1$-$S4 are shown in Table~\ref{tab:tab1}. A comparison of S1 and S2 showed that applying an attention mechanism in conjunction with the original DNN improved deep speaker embedding-based speaker verification performance by 3.2\% EER reduction and 2.3\% $\rm min$$\rm C_{primary}$ reduction. 
These results are consistent with \cite{okabe}.

S3 applied the attention weights from the attentive model to the frame-level features from non-attentive speaker embedding network, and also outperformed S1. 
It also even achieved results comparable to those with S2.
S4 extracted frame-level features from the attentive speaker embedding network and then applied non-attentive pooling by discarding the simultaneously trained attention model. 
Surprisingly, its performance was much worse. 
The comparison of the four systems indicates that: 
(1) attention weights derived from an attentive speaker embedding network can be used with frame-level features from a non-attentive network, and 
(2) decoupling the attention model from an attentive embedding network is detrimental.


\subsection{I-vector extraction with attention weights}
In accord with the results obtained  in Subsection~\ref{ssec:expatt}, we examined a new combination of embeddings and attention weights.
\vspace*{-2mm}
\subsubsection{Experimental settings}

The baseline i-vector system and our proposed system use 20-dimensional MFCCs for every 10ms, the same as with deep speaker embedding systems. 
Their delta and delta-delta features were appended to form 60-dimensional acoustic features. 
Sliding mean normalization with a 3-second window and energy-based VAD were then applied in the same way as was done with deep speaker embedding systems.
An i-vector of 400 dimensions was then extracted from the acoustic feature vectors, using a 2048-mixture UBM and a total variability matrix (TVM). 
Mean subtraction, whitening, and length normalization \cite{garcia2011analysis} were applied to the i-vector in the same way as was done with deep speaker embedding systems, 
and similarity was then evaluated using a PLDA model with a speaker space of 400 dimensions.
For our proposed i-vector extraction with attention weights, the weights were extracted from S2, as described in Subsection~\ref{ssec:expatt}.

\subsubsection{Experimental results and analyses}
Table~\ref{tab:tab2} shows the results of i-vector systems. S5 represents the conventional i-vector baseline.
S6 is the proposed i-vector extraction with the attention weights derived from deep speaker embeddings.
With the proposed method, the i-vector-based system was improved by 6.6\% EER reduction and 3.5\% $\rm min$$\rm C_{primary}$ reduction. 
The attention weights derived from an attention model in a deep speaker embedding network S2 improved not only the matched x-vectors but also the  i-vectors,
which the attention model had never seen. 
This interesting result suggests that the weights from the attention model trained with deep speaker embedding is able to represent the importance regardless of the type of feature representation, i-vector or x-vector.

Note that 
the i-vector and x-vector paradigms have their own advantages under different conditions and with different measures. In this paper, we don't compare across x-vector and i-vector systems.  

\begin{table}[h] 
\centering
\begin{tabular}[t]{l||cc}
\hline
systems & EER(\%) & $\rm min$$\rm C_{primary}$ \\
\hline \hline
S5: baseline &13.04 &0.826 \\
S6: with attention &12.18 &0.797 \\
\hline
\end{tabular}
\caption{Performance of non-attentive and attentive i-vectors}
\label{tab:tab2}
\vspace*{-2mm}
\end{table}
\subsection{Combination of attention mechanisms and soft VAD}

As shown in \cite{yamamoto}, i-vector extraction with voice posteriors as the weights (soft VAD) in extraction improved speaker recognition performance. 
We tried combining attention weights with voice posteriors and replaced the attention weight $\alpha_t$ with $\alpha_t q_t$, the product of the attention weight $\alpha_t$ and voice posterior $q_t$  in Eqs.~\eqref{eq:attmean} and ~\eqref{eq:attsdv} for deep speaker embeddings, and in Eqs.~\eqref{eq:newNc} and ~\eqref{eq:newFc} for i-vector extraction.

\subsubsection{Experimental settings}
We used the same soft VAD reported in \cite{yamamoto}, for which an LSTM (Long Short-Term Memory) neural network was trained for voice posterior estimation.
A subset of the Fisher corpus which consists of only the transcribed segments including noise was utilized as the training data for the VAD.
Five classes were assigned as the LSTM output: voice, laughter, noise, unclear voice, and silence. 
Training was implemented using the nnet3 neural network library in Kaldi's official repository \cite{povey2011kaldi}\cite{snyder2018vector}.
The acoustic features were 40-dimensional MFCCs extracted from a frame of 25ms width at every 10ms.

\subsubsection{Experimental results and analyses}
Results are shown in Table~\ref{tab:tab3}. From S1 and S5, we see that applying the voice posterior as a weight in pooling not only improves the performance in i-vector system (S5) \cite{yamamoto}, but also works in deep speaker embedding system (S1). From S2 and S6, results show that applying soft VAD to an attentive model further improved performance. 

All the experiments described so far were done with systems trained with English telephone recordings from SRE04-10, Switchboard and Fisher. A certain amount of domain difference between the training data and evaluation data exists, including language differences, channel differences. 
In our last experiment, we applied Kaldi's unsupervised domain adaptation \cite{povey2011kaldi} to adapt the PLDA in systems S1$-$S6 using SRE'16 unlabeled development data, which included 2,274 Cantonese and Tagalog utterances.
Here, we refer to pre-adaptation systems as out-of-domain systems and to post-adaptation systems as in-domain systems.

Table~\ref{tab:tab4} shows the performance of adapted in-domain systems, which can be used for comparison with results in Table~\ref{tab:tab3}. 
Adaptation correspondingly improved performance considerably and the trends in performance observed in out-of-domain systems remained the same in in-domain systems. 
We achieved our best results by applying domain adaptation, attention weights,  and soft VAD.

\begin{table}[t] 
\small
\centering
\begin{tabular}[t]{c||rr|rr}
\hline
&\multicolumn{2}{ c }{w/o soft VAD} & \multicolumn{2}{| c }{w/ soft VAD} \\ \cline{2-5}
system & EER(\%) & $\rm minC_{primary}$ & EER(\%) & $\rm minC_{primary}$ \\
\hline \hline
S1 &11.47 &0.873 & 11.02 &0.854\\

S2 &11.10 &0.853 & 10.60 & 0.827\\ 
\hline
S5 &13.04 &0.826 & 12.09 &0.802 \\ 
S6 &12.18 &0.797 & 11.86 & 0.782\\

\hline
\end{tabular}
\caption{Performance of non-attentive and attentive x-vector and i-vector systems with and without soft VAD}
\label{tab:tab3}
\end{table}

\begin{table}[t] 
\small
\centering
\begin{tabular}[t]{c||rr|rr}
\hline
&\multicolumn{2}{ c }{w/o soft VAD} & \multicolumn{2}{| c }{w/ soft VAD} \\ \cline{2-5}
system & EER(\%) & $\rm minC_{primary}$ & EER(\%) & $\rm minC_{primary}$ \\
\hline \hline
S1' &8.39 &0.682 & 8.11 &0.660\\

S2' &8.32 &0.664 & 7.85 & 0.620\\ 
\hline
S5' &11.72 &0.693 & 11.06 &0.669 \\ 

S6' &10.66 &0.667 & 10.58 & 0.656\\
\hline
\end{tabular}
\caption{Performance of non-attentive and attentive x-vector and i-vector systems after domain adaptation}
\label{tab:tab4}
\end{table}

\section{Summary}


This paper has presented an experimental investigation on deep speaker embedding with an attention mechanism.  
Interesting results include  
(1) attention weights derived from an attentive speaker embedding network can be used with frame-level features from a non-attentive network, and 
(2) decoupling the attention model from an attentive embedding network is detrimental.
Inspired by these findings, we have also proposed the application of  attention weights from a deep speaker embedding network to another type of speaker embedding: i-vector. 
Experimental results have shown a 9.0\% EER reduction and a 3.8\% $\rm min$$\rm C_{primary}$ reduction, 
which shows that the attention weights can truly represent the importance of frames regardless of the feature representations of the frames. 
This indicates the possibility of an extension to other speaker embeddings in the future. 
%
Finally, we have shown that combining soft VAD with an attention weight further reduces $\rm min$$\rm C_{primary}$ in deep speaker embedding and i-vector systems, by 6.6\% and 1.6\%, respectively.

\bibliographystyle{IEEEbib}
\bibliography{strings,refs}

\end{document}